\documentclass[aps,physrev,reprint,groupedaddress]{revtex4-2}
\usepackage{amsmath}
\usepackage{graphicx} % Required for inserting images

\usepackage{array}
\usepackage{xcolor}
\usepackage{cancel}
\usepackage[utf8x]{inputenc}
\usepackage[left]{lineno}
\usepackage[normalem]{ulem}
\usepackage{multirow}

\begin{document}

\title{Exploring Fermionic Dark Matter Admixed Neutron Stars in the Light of Astrophysical Observations}

\author{Payaswinee Arvikar \textsuperscript{1,2}}
\email{p20230533@hyderabad.bits-pilani.ac.in}
\author{Sakshi Gautam \textsuperscript{2,3}}
\author{Anagh Venneti \textsuperscript{2}}
\author{Sarmistha Banik \textsuperscript{2}}

\affiliation{\textsuperscript{1}Dharampeth M. P. Deo Memorial Science College, Nagpur 440033, India}
\affiliation{\textsuperscript{2}Department of Physics, BITS-Pilani Hyderabad Campus, Hyderabad, 500078, India}
\affiliation{\textsuperscript{3}Department of Physics, Panjab University, Chandigarh, 160014, India}

\begin{abstract}
We studied the properties of dark matter admixed-neutron stars (DMANS), considering fermionic dark matter (DM) that interacts gravitationally with hadronic matter (HM). Using relativistic mean-field equations of state (EoSs) for both components, we solved the two-fluid Tolman–Oppenheimer–Volkoff (TOV) equations to determine neutron star (NS) properties assuming that DM is confined within the stellar core. For hadronic matter, we employed realistic EoSs derived from low energy nuclear physics experiments, heavy-ion collision data, and NS observations. To constrain key dark matter parameters—such as particle mass, mass fraction, and the coupling to mass ratio— we applied Bayesian inference, incorporating various astrophysical data including mass, radii, and NICER mass-radius distributions for PSR J0740+6620 and PSR J0030+0451. Additionally, we explored the influence of high-density HM EoSs and examined the impact of stiffer hadronic EoSs, excluding the vector meson self-interaction term. Our findings indicate that current astrophysical observations primarily constrain the dark matter fraction, while providing limited constraints on the particle mass or coupling. However, the dark matter fraction is largely insensitive to how astrophysical observations or uncertainties in the high-density EoS are incorporated. Instead, it is predominantly determined by the stiffness of the hadronic EoS at high densities, with stiffer hadronic EoSs yielding a higher dark matter mass fraction. Therefore, we conclude that the dark matter fraction plays a  crucial role in shaping the properties of DMANS. Future investigations incorporating more realistic EoSs and astrophysical observations of other compact objects may provide deeper insights into dark matter.
\end{abstract}

\maketitle

\section{Introduction}
Dark matter (DM), an enigmatic component of the universe comprising approximately 27\% of its mass-energy content, remains one of the most profound mysteries in astrophysics and cosmology. Unlike ordinary matter, it does not interact via electromagnetic forces, rendering it undetectable through conventional observational techniques such as light or radiation. Instead, its existence is inferred from its gravitational influence on galaxies, galaxy clusters, and large-scale cosmic structures \cite{GOLOVKO2023106164, rotationcurve_Ludwig,Turner_2000,Del_Popolo_2007}. Despite decades of extensive research, the fundamental properties and nature of DM continue to elude definitive understanding, posing one of the most pressing challenges in modern physics.

The quest to unravel the true nature of DM has led researchers to propose a diverse array of potential dark matter candidates. These include weakly interacting massive particles (WIMPs) \cite{Kouvaris2011,WIMP_Bertone_2018}, feebly interacting massive particles (FIMPs) \cite{Hall_2010,FIMP_Bernal_2017,DSen_2021}, axions \cite{Axion_Duffy_2009}, sterile neutrinos \cite{DMreview_Bertone_2018}, self-interacting DM \cite{Xiang_2014,Ivanytskyi_2020,Thakur_2024}, and non-gravitationally interacting DM \cite{Panotopoulos_2017,Das_2019,Sagun_2022,Routaray_2023,AnkitKumar_2024}, bosonic dark matter \cite{Karkevandi2021,Konstantinou_2024,Buras-Stubbs_2024}, fermionic DM \cite{Li_2012,Ivanytskyi_2020,thakur2023exploring} among others, with mass ranging from a few eVs to GeVs. Despite decades of investigation, 
no definitive evidence has yet identified the exact nature of DM.  
Concurrently, a variety of experimental and observational tools have been developed to probe its existence. Direct detection experiments, such as XENON100 \cite{Aprile_2012} and PANDAX-II \cite{Wang_2020}, aim to observe DM interactions within highly sensitive detectors, while space-based observatories like the James Webb Space Telescope seek to explore large-scale cosmic structures \cite{JWST},  
potentially uncovering indirect signatures of DM, among numerous others indirect means \cite{DONATO2014,Heros2020,Biondini:2023}. 

Neutron stars (NS), the ultra-dense remnants of massive stars that have undergone supernova explosions, serve as  a unique laboratory for  exploring potential interactions between dark matter and ordinary matter under extreme conditions.
DM can be accumulated into NSs through various mechanisms operating over its lifetime, from protostar to NS phase. With  spherically symmetric accretion, the accumulated DM mass for a NS of canonical mass may range from approximately $10^{-13}$ M$_{\odot}$ in regions near the solar neighborhood to $10^{-5}$ M$_{\odot}$ in the vicinity of the Galactic center, depending on the local DM density \cite{DelPopolo:2019,Rutherford_2023}. Additionally, DM production may occur via baryonic processes during core-collapse supernovae or in the aftermath of binary neutron star mergers. However, these standard accretion and production channels typically yield relatively low DM content, with the DM fraction within the NS remaining at the level of a few percent. Conversely, there are arguments like NS residing in DM clumps or near the galactic center may accumulate larger fractions \cite{DelPopolo:2019,Deliyergiyev:2023}.

Recent theoretical studies suggest that DM may accumulate inside NS, influencing their internal structure and potentially leading to observable effects on their mass, radius, and cooling rates \cite{Kouvaris_2010,Ivanytskyi_2020, Avila_2023,Giangrandi_2024}. The interaction between DM and baryonic matter in these extreme environments offers valuable insights into key DM properties, including particle mass, interaction cross-section, and possible coupling mechanisms \cite{Lavallaz_2010, Nelson_2019, Ivanytskyi_2020, Karkevandi2021, Miao_2022,HCDas_2022,Collier_2022, Leung_2022, Guha:2024}. Given the elusive nature of DM, both fermionic and bosonic candidates have been explored to investigate the properties of dark matter-admixed neutron stars (DMANS) \cite{Ellis_2018,Das_2019, Ivanytskyi_2020, Kain_2021, Routaray_2023, thakur2023exploring, Giangrandi_2024,Li_2012,thakur2023exploring,Guha_2021,Husain_2021,Rutherford_2023}.

For instance, Li {\it et al.} \cite{Li_2012} investigated fermionic DM particles and established an upper limit of approximately 0.64 GeV on their mass to account for the 2M$_\odot$ NS observation of PSR J1614-2230. Similarly, the effect of DM fermion mass on the properties of both static and rotating NSs was examined in \cite{Guha_2021}, showing that as the DM particle mass increases, the maximum gravitational mass, and radius and tidal deformability of a 1.4M$_\odot$ NS decrease. Another study  \cite{thakur2023exploring} further explored the correlations between fermionic DM model parameters and NS properties, incorporating uncertainties in the nuclear sector.
In contrast, Husain {\it et al}. \cite{Husain_2021} analyzed NS properties under the assumption of bosonic DM, finding that increasing DM content leads to a reduction in both the maximum NS mass and the tidal deformability of a 1.4M$_\odot$ NS. Similarly, Rutherford {\it et al.} \cite{Rutherford_2023} attempted to constrain bosonic asymmetric DM parameters using NS mass-radius measurements, but concluded that current uncertainties in the baryonic equation of state (EoS) hinder precise constraints on DM properties. The impact of DM on NS oscillations, including {\it f-mode} and {\it r-mode} oscillations, has also been explored in various studies \cite{Shirke23, HCdas2023, flores2024, dey2024,ThakurPratik2024}. 

Notably, different studies suggest the variation of permissible DM fractions in NSs. For instance, Ref. \cite{thakur2023exploring} estimated a DM mass fraction of approximately 10–25\% when applying a 1.9M$_\odot$ NS constraint. However, Ref. \cite{Karkevandi2021} favored a sub-GeV DM particle with a DM mass fraction of around 5\%, based on the observation of 2.0M$_\odot$ NS and constraints of tidal deformability from the LIGO/Virgo collaboration. Similarly, Ref. \cite{Routaray_2023} established an upper limit of approximately 10\% mass fraction for DM in NSs, incorporating NICER observations alongside mass-radius constraints from HESS J1731-347. Furthermore, recent studies, such as \cite{Guha:2024}, have restricted the mass of the DM particle to a range of 0.1 to 30 GeV using NICER data and the GW170817 event. Ref. \cite{Ivanytskyi_2020} further placed an upper bound of approximately 60 GeV on the DM particle mass, incorporating additional observational constraints. These findings underscore the ongoing uncertainty regarding both the DM fraction in NSs and the mass of DM particles, highlighting the need for further theoretical refinements, consistent with current observational constraints on compact stars. 

In this study, we aim to constrain the parameter space of the fermionic DM using Bayesian analysis, informed by astrophysical observations of mass, radius, and tidal deformability of NS. Additionally, we investigate how the treatment of these observations affects dark sector parameters. Our approach employs recently developed nuclear matter EoSs \cite{Venneti_2024,SGAUTAM_2024}, which incorporate constraints from finite nuclei properties, supplemented by experimental data from heavy-ion collisions (HIC) and astrophysical observations of compact stars. Finite nuclei constraints (FNC) to the EoSs are applied explicitly—by enforcing binding energy and charge radius constraints for select nuclei—and implicitly, where constraints obtained from nuclear masses are imposed. Further constraints on symmetry energy, symmetric nuclear matter pressure, symmetry pressure, and incompressibility are derived from HIC experiments and studies on finite nuclei. At higher densities, astrophysical observations of NS mass, radius, and tidal deformability provide additional constraints on the EoSs. Furthermore, we consider pure gravitational interactions between hadronic and DM components to explore the properties of DMANS. The parameters of the DM model are constrained using various astrophysical data, including mass, radii, and NICER mass-radius distributions for PSR J0740+6620 and PSR J0030+0451. These observations are integrated into a statistical inference framework to explore the DM parameter space and evaluate their impact on the properties of DMANS.

Note that when DM interacts gravitationally with hadronic matter, it can accumulate inside a NS under the following conditions— either with the hadronic matter having a smaller radius than the DM distribution, or vice versa. In the former scenario, DM is concentrated in the core of the NS, while in the latter scenario, it forms an extended halo around it. The transition between these configurations is known to depend on factors such as the mass of the DM particle, the coupling strength, and the DM fraction \cite{Miao_2022, AnkitKumar_2024,Shawqi:2024}. For instance, DM particles with masses on the order of a few hundred MeV typically form an extended halo around a neutron star, whereas more massive DM particles tend to be gravitationally confined to the stellar core \cite{Ivanytskyi_2020,Giangrandi_2024}. In this study, however, we assume that DM is entirely confined to the neutron star's core.

The paper is structured as follows. Section \ref{EoS HM DM} provides a brief overview of the models for hadronic and DM EoSs, along with high-density refinements in the EoS for the hadronic sector. Section \ref{bayesian} outlines the various scenarios considered for Bayesian analysis. The results are presented and discussed in Section \ref{results}. Finally, we summarize our findings in Section \ref{summary}.

\section{Equation of state for nuclear matter and dark matter}\label{EoS HM DM}
\subsection{Hadronic matter model}
We consider here the Relativistic Mean Field (RMF) approach for the hadronic matter (HM), composed of neutrons and protons interacting through the exchange of three mesons, $\sigma, \omega$ and $\rho$. The isoscalar-scalar meson $\sigma$ provides medium-range attractions between nucleons, isoscalar-vector meson $\omega$ provides short-range repulsion between nucleons, and isovector-vector meson $\rho$ accounts for the isospin asymmetry. 

The Lagrangian density for HM is given by \cite{Venneti_2024,Dutra_2014},

 \begin{equation*} %\label{lag}
   \mathcal{L}_{NL} = \mathcal{L}_{nm} + \mathcal{L}_{\sigma} + \mathcal{L}_{\omega} + \mathcal{L}_{\rho} + \mathcal{L}_{int},
\end{equation*}
where,
\begin{eqnarray}\label{eq:lag}
    \mathcal{L}_{nm}&=& \!\bar{\psi}\left(i\gamma^\mu\partial_\mu - M\right)\psi +\!\!g_\sigma \sigma \bar{\psi} \psi -\!\!g_\omega \bar{\psi}\gamma^\mu\omega_\mu\psi 
    \nonumber \\
     &&-\frac{g_\rho}{2}\bar{\psi}\gamma^\mu\vec{\rho}_\mu \vec{\tau}\psi,\nonumber \\
%\begin{eqnarray}
    \mathcal{L}_\sigma&=&\frac{1}{2}\left(\partial^\mu\sigma\partial_\mu\sigma - m^2_\sigma \sigma^2\right) - \frac{\text{A}}{3} \sigma^3 - \frac{\text{B}}{4} \sigma^4 \nonumber,\\
%\end{eqnarray}
%\begin{eqnarray}
    \mathcal{L}_\omega&=&\! -\frac{1}{4} \Omega^{\mu\nu}\Omega_{\mu\nu} + \frac{1}{2}m^2_\omega\omega^\mu\omega_\mu +\frac{\text{C}}{4}\left(g_\omega^2 \omega_\mu \omega^\mu\right)^2 \nonumber,\\
%\end{eqnarray}
%\begin{eqnarray}
    \mathcal{L}_\rho&=& -\frac{1}{4}\vec{B}^{\mu\nu}\vec{B}_{\mu\nu} + \frac{1}{2}m^2_\rho \vec{\rho}_\mu \vec{\rho}^{ \mu} , \nonumber \\
%\end{eqnarray}
%and
%\begin{eqnarray}
    \mathcal{L}_{int}&=&\frac{1}{2} \Lambda_v g^2_\omega g^2_\rho \omega_\mu \omega^\mu \vec{\rho}_\mu \vec{\rho}^\mu. 
\end{eqnarray}
In the above equations, $\Omega_{\mu\nu} = \partial_\nu \omega_\mu - \partial_\mu \omega_\nu$ and $\vec{B}_{\mu\nu} = \partial_\nu \vec{\rho}_\mu - \partial_\mu \vec{\rho}_\nu - g_\rho \left(\vec{\rho}_\mu \times \vec{\rho}_\nu \right)$.  
Here, $\psi$ is the wave function for the nucleons of mass M. The $\sigma$, $\omega$ and $\rho$ mesons have masses and couplings denoted by $m_\sigma$, $m_\omega$, $m_\rho$ and $g_\sigma$, $g_\omega$, $g_\rho$ respectively. In addition, there are other self-interactions and cross-interactions between scalar and vector mesons, whose coupling strengths are represented by $A, B, C$ and $\Lambda_{v}$. Note that these couplings are calibrated to yield the measured values of the bulk properties of finite nuclei as well as neutron star observables accurately. Based on the above Lagrangian, the energy density $\mathcal{E}$ and pressure $P$ of hadronic matter can be computed at a given number density from the following equations \cite{Venneti_2024},

\begin{widetext}
\begin{eqnarray}{\label{eos_nm}}
    \mathcal{E}_{HM} = \frac{1}{\pi^2} \int_{0}^{k_p,k_n} dk k^2 \sqrt{k^2 + (M^*)^2}  + \frac{1}{2} m_\sigma^2 \sigma^2 - \frac{1}{2} m_\omega^2 \omega^2 -  \frac{1}{2} m_\rho^2 \rho^2 + \frac{\text{A}}{3} \sigma^3 + \frac{\text{B}}{4} \sigma^4 \nonumber \\ \nonumber
    - \frac{\text{C}}{4} g_\omega^4 \omega^4 + g_\omega \omega(\rho_p + \rho_n) - \frac{\Lambda_v}{2} (g_\rho g_\omega \rho \omega )^2 + \frac{g_\rho}{2} \rho (\rho_p - \rho_n), \\ \nonumber
    P_{HM} = \frac{1}{3 \pi^2} \int_{0}^{k_p,k_n} dk \frac{k^4} {\sqrt{k^2 + (M^*)^2}} - \frac{1}{2} m_\sigma^2 \sigma^2 + \frac{1}{2} m_\omega^2 \omega^2 +  \frac{1}{2} m_\rho^2 \rho^2 - \frac{\text{A}}{3} \sigma^3 - \frac{\text{B}}{4} \sigma^4 \\ %\nonumber
    + \frac{\text{C}}{4} g_\omega^4 \omega^4 +  \frac{\Lambda_v}{2} (g_\rho g_\omega \rho \omega )^2. \\ \nonumber
\end{eqnarray}
\end{widetext}

In the above equations, $\rho_{p,n}$ and $k_{p,n}$ represent the density and the Fermi momentum of protons or neutrons and $M^{*}=M-g_{\sigma}\sigma $ represents the effective nucleonic mass.

We have constrained the distributions of EoSs, derived from the RMF model, under two distinct approaches of implementing finite nuclei constraints. In the first approach, we explicitly incorporate precisely measured binding energies and charge radii of $^{40} Ca$ and $^{208}Pb$ nuclei, along with experimental data from heavy-ion collisions, other finite nuclei observables, and astrophysical observations of neutron stars \cite{Tsang2024} to establish posterior EoS distributions. The second, more commonly used approach implicitly includes FNC through specific properties of symmetric nuclear matter and the density dependence of symmetry energy at sub-saturation densities, in place of constraints obtained from bindings and chare radii. Other constraints obtained from studies of HIC and finite nuclei and astrophysical observations of compact stars are retained unchanged. In this work, we utilize the EoS distributions, obtained after explicit and implicit treatment of FNC and refer to them as BITSH-E and BITSH-I respectively. Furthermore, we set the vector self-interaction term to be zero (C=0) in the $\mathcal{L}_{\omega}$ of Eq. (1) to examine our results for stiffer EoSs. The coupling parameters used for these two models with and without the inclusion of vector self-interactions are listed in Table \ref{table:couplings}. The EoS of the core is obtained from Eq. \ref{eos_nm}. The crust of neutron stars is not very well known, and various approaches have been followed to model this part of the neutron star. Here, we use the EoS for outer crust by Baym-Pethick-Sutherland (BPS) \cite{Baym:1971}  till a density of $0.0016 \rho_0$, whereas, the inner crust is characterized as a polytropic fit \cite{Malik_poly:2017}. connected to the core EoS. The EoS of the core is assumed  to start at 0.5$\rho_0$, which is obtained for the Lagrangian given by Eq. 1.

\begin{table*}[t]
    \centering
    \caption{Coupling constants of the Lagrangian(Eq. \ref{eq:lag}) used for the BITSH-E and BITSH-I EoSs. Masses (in MeV) of nucleon and mesons; $M$, $m_\sigma$, $m_\omega$ , $m_\rho$ are 939, 508.1941, 782.501 and 763, respectively. All the couplings below are dimensionless except $A$ which is in $fm^{-1}$. } 
    \renewcommand{\arraystretch}{1.2}
    %\begin{tabular}{|l|l |c |c |c |c |c |c |c|}
    \begin{tabular}{|*{2}{>{\centering\arraybackslash}p{0.1\linewidth}|} *{7}{>{\centering\arraybackslash}p{0.09\linewidth}|}}
    \hline\hline 
    \centering
    %\vspace{0.1cm}
	\small{Model} & Interaction & $g_\sigma$ & $g_\omega$ &  $g_\rho$ & $A$ & $B$ & C & $\Lambda_v$     \\
    \hline\hline 
    %\vspace{0.1cm} 
	\small{BITSH-E} & \small{with $\omega^4$ \hspace{0.4cm} (C $\ne$ 0)} & 10.0 & 12.52 & 10.32 & 10.62 & -14.07 & 0.0027 & 0.08 \\
    \hline    
    %\vspace{0.1 cm}
    \small{BITSH-I} & \small{with $\omega^4$ \hspace{0.4cm} (C $\ne$ 0)} & 9.05 & 10.69 & 9.95 & 15.29 & -13.56 & 0.0006 & 0.06 \\
    \hline\hline 
    %\vspace{0.1 cm}
    \small{BITSH-E} & \small{without $\omega^4$ (C $=$ 0)} & 9.58 & 11.58 & 10.44 & 15.13 & -38.21 & - & 0.1702 \\ 
    \hline 
    %\vspace{0.1cm}
    \small{BITSH-I} & \small{without $\omega^4$ (C $=$ 0)} & 8.96 & 10.35 & 9.95 & 19.71 & -46.44 & - & 0.0726\\
    \hline\hline 
	\end{tabular}
    \label{table:couplings}
\end{table*}

\subsection{Dark matter model}\label{DM_EoS}

Numerous ways are reported in the literature to incorporate dark matter, {\it viz}. fermionic or bosonic, self-interacting (via attractive or repulsive interactions) or even non-interacting. For DM, here we consider a RMF approach where DM particles ($\chi_D$) are considered to be fermions with mass $M_D$. A `dark vector meson' ($V_D^\mu$) of mass $m_{vd}$ couples to DM particle via coupling strength $g_{vd}$. The Lagrangian density for the DM sector is given as,

\begin{widetext}
\begin{equation}
    \mathcal{L}_\chi = \!\bar{\chi}_D [\gamma_\mu (i\partial^\mu - g_{vd} V_D^\mu ) - M_D ] \chi_D - \frac{1}{4} V_{\mu\nu,D} V_D^{\mu\nu} + \frac{1}{2} m^2_{vd} V_{\mu,D} V_D^\mu.
\end{equation}
\end{widetext}

The equations for energy density and pressure for DM are given as \cite{Thakur_2024},
\begin{eqnarray}{\label{eos_dm1}}
    \mathcal{E}_{DM} &=& \frac{1}{\pi^2} \int_{0}^{k} dk k^2 \sqrt{k^2 + (M_D)^2} + \frac{g_{vd}^2}{2m_{vd}^2} \rho_D^2, \nonumber \\  %+ \frac{m_{sd}^2}{2g_{sd}^2} (M_D - M_D^*)^2 \\ 
    P_{DM} &=& \frac{1}{3 \pi^2} \int_{0}^{k} dk \frac{k^4} {\sqrt{k^2 + (M_D)^2}} + \frac{g_{vd}^2}{2m_{vd}^2} \rho_D^2 ,%- \frac{m_{sd}^2}{2g_{sd}^2} (M_D - M_D^*)^2
\end{eqnarray}
where $k$ is the Fermi momentum for DM. The ratio $C_{vd} = g_{vd}/m_{vd}$ is for vector interaction between DM particles and dark mesons that play a crucial role in modeling of DM EoS and $\rho_D$ represents the density of DM fermions associated with the mean field value of dark vector mesons.

\subsection{Interaction between hadronic matter and dark matter: two-fluid formalism}

We consider HM and DM to interact solely through gravitational interaction. Therefore, the Lagrangians for the two fluids are independent and energy-momentum tensors for each fluid are conserved separately. With the EoS equations (Eqs. \ref{eos_nm} and \ref{eos_dm1}) of an electrically neutral, relativistic free Fermi gas of HM and DM in chemical equilibrium, the mass and radius of NS is calculated solving the Tolman-Oppenheimer-Volkoff (TOV) equations for the two-fluid system given by,

\begin{eqnarray}{\label{tov}}
    \frac{dP_{HM}}{dr} &=& -(P_{HM}+\mathcal{E}_{HM}) \frac{4 \pi r^3 (P_{HM}+P_{DM}) + m(r)}{r(r-2m(r))} ,\nonumber\\ 
    \frac{dP_{DM}}{dr} &=& -(P_{DM}+\mathcal{E}_{DM}) \frac{4 \pi r^3 (P_{HM}+P_{DM}) + m(r)}{r(r-2m(r))} ,\nonumber \\ 
    \frac{dm (r)}{dr} &=& 4 \pi (\mathcal{E}_{HM}(r) + \mathcal{E}_{DM}(r)) r^2 .
\end{eqnarray}
where $m(r)$ is the mass enclosed in radius $r$. To control the amount of dark matter present in DM admixed NS, one can change the ratio of central energy density of DM to that of HM; ie. $f_{c} = \mathcal{E}_{DM}/\mathcal{E}_{HM}$ \cite{das2020dark, Xiang_2014} so that the ratio of DM mass to the total mass of NS; $f_{DM}= \frac{M_{DM}}{M_{Total}}$ remains fixed \cite{ Sagun_2022,Routaray_2023, thakur2023exploring,Thakur_2024}. 

$M_{DM}$ is the mass enclosed by a radius $R_{DM}$, given by 

$M_{DM}(R_{DM}) = 4 \pi \int_0^{R_{DM}} r^2 \mathcal{E}_{DM}(r) dr$. The total mass of the star is calculated by $M_{Total}(R) = 4 \pi \int_0^{R} r^2 (\mathcal{E}_{HM}(r)+\mathcal{E}_{DM}(r)) dr$, where $R$ is the radius of the star where the pressure vanishes.  Note that in the present study, we use $f_{DM}$ to quantify the amount of DM fraction in DMANS and the values of $f_{DM}$ are listed in percentage.

In a binary NS system, during the final stages of inspiral, the tidal gravitational field generated by a companion causes the two neutron stars to undergo quadrupole deformations.  Due to this, tidal forces are exerted and the magnitude of deformation that occurs is described as tidal deformability, which is quantified as,
$\lambda = \frac{2}{3} k_2 R^5$ and the dimensionless tidal deformability is given by $\Lambda = \frac{2}{3} k_2 \mathcal{C}^{-5}$. Here compactness $\mathcal{C} = M/R$ and $k_2$ is the tidal love number which is given by,

\begin{widetext}
\begin{multline}{\label{k2}}
    k_2 = \frac{8\mathcal{C}^5}{5} (1-2\mathcal{C})^2 \bigl[2+2\mathcal{C}(y_R-1) - y_R \bigr] \times \Bigl( 2\mathcal{C} (6 - 3y_R + 3\mathcal{C}(5 y_R - 8)) \\ 
    + 4\mathcal{C}^3 \bigl[13 - 11 y_R + \mathcal{C}(3y_R - 2) + 2\mathcal{C}^2(1+y_R) \bigr] \\
    + 3(1 - 2\mathcal{C})^2 \bigl( 2 - y_R + 2\mathcal{C}(y_R-1)\bigr) log(1-2\mathcal{C}) \Bigr) ^{-1}.
\end{multline}

Here $y_R$ is the auxiliary variable obtained by solving the following differential equation

\begin{equation}
    r\frac{dy(r)}{dr} + y^2(r) + y(r)F(r) + r^2Q(r) = 0 .
\end{equation}

Here, $F(r)$ and $Q(r)$ for a two-fluid system are given by \cite{thakur2023exploring},

\begin{equation}
    F(r) = \frac{r-4\pi r^3(\mathcal{E}_{HM}(r)+\mathcal{E}_{DM}(r))-(P_{HM}(r)+P_{DM}(r))}{r-2m(r)},
\end{equation}

and 

\begin{multline}
    Q(r) = \frac{4 \pi r \biggl(5(\mathcal{E}_{HM}(r)+\mathcal{E}_{DM}(r))+9(P_{HM}(r)+P_{DM}(r))+\frac{\mathcal{E}_{HM}(r)+P_{HM}(r)}{\partial P_{HM}(r)/\partial \mathcal{E}_{HM}(r)}+ \frac{\mathcal{E}_{DM}(r)+P_{DM}(r)}{\partial P_{DM}(r)/\partial \mathcal{E}_{DM}(r)} - \frac{6}{4\pi r^2}\biggr)}{r-2m(r)}  \\
    - 4 \biggl[\frac{m(r)+4\pi r^3(P_{HM}(r)+P_{DM}(r))}{r^2 (1-2m(r)/r)}\biggr]^2.
\end{multline}
\end{widetext}

\subsection  {High density NS EoS}\label{HD EoS}

The core of the NS is not fully understood and an ambiguity in its internal composition could lead to an emergence of new degrees of freedom such as quarks, hyperons or kaons at higher densities. This uncertainty in EoS at high densities ($\rho>2\rho_{0}$) is often incorporated through speed of sound parametrization. The high-density part of the EoS joins smoothly to the low density part ($\rho\leq2\rho_{0}$) such that the velocity of the sound never exceeds the velocity of light and asymptotically approaches the conformal limit. The speed of sound in the higher density region is given by \cite{Tews_2018},
\begin{widetext}
\begin{equation}\label{cs2}
    c_s^2 = \frac{1}{3} -c_1 \text{exp}\bigl[ -\frac{(n-c_2)^2}{n_{bl}^2}\bigr] + h_p \text{exp}\bigl[\frac{(n-n_p)^2}{w_p^2}\bigr] \bigl(1+\text{erf}[s_p\frac{(n-n_p)}{w_p}]\bigr).
\end{equation}
\end{widetext}
Here, the parameters $h_p$ and $n_p$ represent the maximum speed of sound and the density around which it happens. The parameters $w_p, n_{bl}$ control the width of the curve and $s_p$ is the shape or skewness parameter. The parameters $c_1$ and $c_2$ are determined by the continuity of the speed of sound and its derivative at the interface of low and high density (2$\rho_0$). 
These parameters are allowed to vary within the priors as discussed in \cite{Tews_2018} to add uncertainty to the HM EoS while performing Bayesian analyses.

\section{Bayesian Interface with the Astrophysical Observations}\label{bayesian}

We employ a Bayesian statistical inference framework for the estimation of the DM parameters. Bayes' theorem \cite{stuart1994kendalls}, %\SG{add ref. here}, 
\begin{equation}\label{bayes}
    P(\mathbf{\Theta}|D) = \frac{\mathcal{L}(D|\mathbf{\Theta})P(\mathbf{\Theta})}{\mathcal{Z}},
\end{equation}
connects the conditional probability of the prior $P(\mathbf{\Theta})$ to the posterior $P(\mathbf{\Theta}|D)$ through the likelihood $\mathcal{L}(D|\mathbf{\Theta})$ for a given set of parameters $\mathbf{\Theta} = \{C_{vd}, M_D, f_{DM}\}$ and the astrophysical observations D.

We constrain the parameters of the DM model $C_{vd}$, $M_D$ and $f_{DM}$ by incorporating constraints coming from astrophysical observations. The observation of gravitational waves coming from the binary NS merger event GW170817 \cite{Abbott_2018} places limits on the tidal deformability of the canonical mass (1.4$M_\odot$) NS. This gravitational wave data is incorporated into $\mathcal{L}_{GW}(D|\mathbf{\Theta})$. In addition to the GW observations, we explore different methods of imposing the constraints of NS mass-radius coming from NICER. These observations can be incorporated into the likelihood $\mathcal{L}_{Obs}(D|\mathbf{\Theta})$, and the total likelihood $\mathcal{L}(D|\mathbf{\Theta})$ is defined as, 
\begin{equation}
    \mathcal{L}(D|\mathbf{\Theta}) =  \mathcal{L}_{Obs}(D|\mathbf{\Theta}) \times \mathcal{L}_{GW}(D|\mathbf{\Theta}).
\end{equation}

We analyse the influence of imposing the mass-radius constraints of NS into the likelihood in the following ways.

\vspace{0.5cm}
\textit{\textbf{Case I:}}
The direct mass measurements of NS masses with the help of the Shapiro delay in NS binary systems provide us with a lower bound on the maximum mass of a NS.
As a first case, %\AV{CASE I write up as a paragraph}
the NS maximum mass constraint is applied as a stringent cut at 2.073 $\pm$ 0.069 $M_{\odot}$ \cite{Salmi_2024} for the observations of pulsar PSR J0740+6620. The radius measurements for the pulsar PSR J0030+0451; $R_{1.34} = 12.71 ^{+1.14}_{-1.19}$ km \cite{Riley_2019} (incorporated as $\mathcal{L}_{R_{1.34}}$) and $R_{1.44} = 13.02 ^{+1.24}_{-1.06}$ km \cite{Miller_2019} (incorporated as $\mathcal{L}_{R_{1.44}}$) are imposed using the following sigmoid function to calculate the likelihood.

\begin{equation}
    \mathcal{L}_{\sigma}(D|\mathbf{\Theta}) = \frac{1}{\sqrt{2 \pi \sigma^2}} \exp \biggl ( -\frac{(D(\mathbf{\Theta})-D_{Obs})^2}{2 \sigma^2}\biggr).
\end{equation}
Also, $\mathcal{L}_{GW}(D|\mathbf{\Theta})$ is calculated with GW170817 observation $\Lambda_{1.4} = 190^{+390}_{-120}$ \cite{Abbott_2018} using a similar function. All data are assumed to follow a symmetric Gaussian distribution and the likelihood $\mathcal{L}_{Obs}(D|\mathbf{\Theta}) = \mathcal{L}_{R_{1.34}} \times \mathcal{L}_{R_{1.44}}$.

\vspace{0.5cm}
\textit{\textbf{Case II:}}
Furthermore, simultaneous mass-radius measurements using X-ray hot spots on the NS surface from the NICER mission provide us with additional constraints to be placed on the NS properties. We place these constraints in likelihood function $\mathcal{L}_{NICER}(D|\mathbf{\Theta})$.
In {\it Case II}, we use Kernel Density Estimator (KDE) in our likelihood to use the entire NICER and GW posterior datasets available. Here, the NICER data for the observations of PSR J0030+0451 \cite{Riley_2019,Miller_2019}(low mass) and PSR J0740+6620 \cite{Riley_2021,Miller_2021} (high mass) are imposed to calculate $\mathcal{L}_{Obs}(D|\mathbf{\Theta})$.  The NICER data for PSR J0740+6620 (PSR J0030+0451) is incorporated in the likelihood function as $\mathcal{L}_{NICER}^{high}$($\mathcal{L}_{NICER}^{low}$) and the total likelihood $\mathcal{L}_{Obs}(D|\mathbf{\Theta})$ becomes,
\begin{equation}
\mathcal{L}_{Obs}(D|\mathbf{\Theta}) = \mathcal{L}_{NICER}^{high} \times \mathcal{L}_{NICER}^{low}\nonumber,
\end{equation}
where,
\begin{equation}
    \mathcal{L}_{NICER}(D|\mathbf{\Theta}) = \int^{M_{max}}_{M_0} dm P(m|\mathbf{\Theta}) \times P(D|m,R(m,\mathbf{\Theta})).
\end{equation}
Using this probability, we calculate the likelihood for both the pulsar observations. Also, for GW observation of binary system, $m_1, m_2$ are the masses and $\Lambda_1, \Lambda_2$ are the tidal deformabilities of the binary components.
\begin{multline}
    \mathcal{L}_{GW}(D|\mathbf{\Theta}) = \int^{M_u}_{m_l}dm_1 \int^{m_1}_{M_l} dm_2 P(m_1 , m_2|\mathbf{\Theta}) \\ 
    \times P (d_{GW} |m_1 , m_2 , \Lambda_1 (m_1 ,\mathbf{\Theta}), \Lambda_2 (m_2 , \mathbf{\Theta})).
\end{multline}
This gives the likelihood $L_{GW}$, where $P(m|\Theta)$ is written as,

\begin{equation}
    P(m|\mathbf{\Theta}) = \begin{cases}
        \frac{1}{M_{max}-M_0}  & \text{if}\;  M_0 \leq m \leq M_{max} \\ \nonumber
        0 & \text{else}
    \end{cases}
\end{equation}
For these calculations, $M_0$ is always taken to be 1M$_{\odot}$ and $M_{max}$ is the maximum mass of the neutron star for given set of parameters.

\vspace{0.5cm}
\textit{\textbf{Case III:}}
Next-generation detectors and observatories, such as the Einstein Telescope and Cosmic Explorer \cite{Evans:2023euw,LIGO_2017,Punturo_2010, ET:2019dnz}, are expected to enable more precise measurements of NS through gravitational waves and other astrophysical observations. This will further refine the constraints on NS properties, enhancing our present understanding of compact stars.
In anticipation of  improved precision in the mass and radius measurements of the pulsars PSR J0030+0451 and PSR J0740+6620 from such future observations, we construct a mock data set that simulates current measurements, but with reduced uncertainties. The kernel density estimates (KDEs) generated for this mock data are then incorporated into the likelihood analysis in a similar fashion as in the previous case II. 

We used the Bayesian sampler PyMultiNest \cite{Pymultinest_Buchner_2014} based on the nested sampling algorithm. The following uniform priors for the parameters of the DM model; $C_{vd}$  = 0.0005MeV$^{-1}$ - 0.030MeV$^{-1}$, $M_D$= 500MeV - 3000MeV  and $f_{DM}$ = 0 - 20\% \cite{Xiang_2014, Karkevandi2021,
Routaray_2023, Guha:2024, thakur2023exploring} are used.  The posterior distributions derived from the statistical analyses are presented in Table \ref{table:posteriors}.

\begin{figure*}[t]
    \centering
    \includegraphics[width=0.52\linewidth]{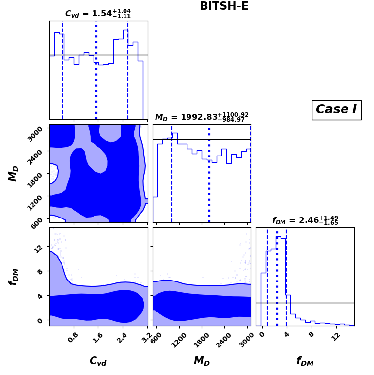}\includegraphics[width=0.52\linewidth]{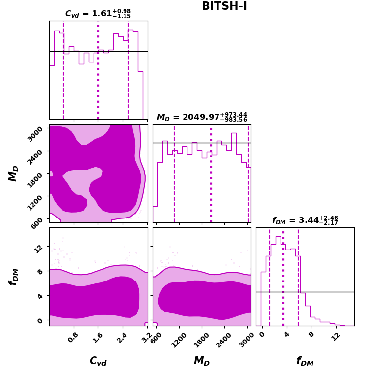}
    \caption{Posterior distributions of DM parameters for BITSH-E (left) and BITSH-I (right). HM EoS is fixed at all densities. Astrophysical observations are treated as per {\it Case I}. The 1$\sigma$ CI is displayed as vertical dashed lines in the marginalized posterior distributions of DM parameters. The dotted line represents the median for the distributions. $C_{vd}$ is mentioned in units MeV$\times10^{-2}$, $M_D$ in MeV and $f_{DM}$ is in percentage. The black horizontal lines in the diagonal panels represent the prior distributions.}
    \label{fig:corner_case1_1}
\end{figure*}

\section{Results and Discussion}\label{results}

We modeled the dense matter in DMANS using a fermionic DM EoS and HM EoS, incorporating constraints from finite nuclei, heavy-ion collisions, and astrophysical observations. DM and HM are interacting gravitationally only and TOV equations are solved using two-fluid formalism.  Bayesian inference is performed to obtain posterior distributions of the parameters of the dark sector using different ways to incorporate astrophysical observations as discussed in the previous section.
In {\it Case I}: only constraints of maximum mass from the pulsar PSR J0740+6620 and radii of pulsar PSR J0030+0451 at low masses are implemented.\\
{\it Case II}: the full NICER mass-radius data sets for these pulsars are used.\\
{\it Case III}: a mock data set generated with reduced uncertainty is imposed as constraints.
Additionally, we investigate the role of uncertainty in HM EoS at high densities. To assess the robustness of our findings, we repeat  the analysis for comparatively stiffer HM EoSs (where self-interaction term of vector mesons is omitted, C = 0) to  evaluate the influence of HM EoS on the modeling the DM parameters.

\subsection{ Constraining DM parameters for HM EoSs fixed over a range of densities} \label{onlyHM}

First, we present the corner plots that illustrate the marginalized posterior distributions
%Fig. \ref{fig:corner_case1_1} shows 
for the BITSH-E (left) and BITSH-I (right) models (with C $\neq$ 0). The diagonals of the corner plots show the marginalized posterior distributions for each DM parameter, with vertical dashed lines indicating the $1 \sigma$ confidence interval (CI). The off-diagonal plots illustrate the pairwise probability distributions for the three DM parameters, with contours enclosing the $1$ and $2 \sigma$ CIs.

\begin{figure*}[t]
    \centering
    \includegraphics[width=0.52\linewidth]{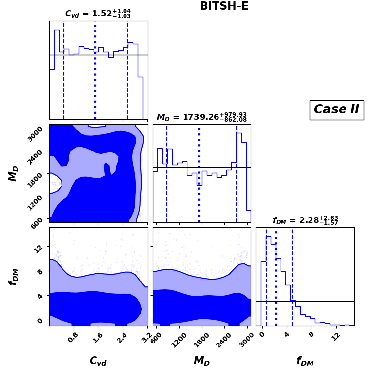}\includegraphics[width=0.52\linewidth]{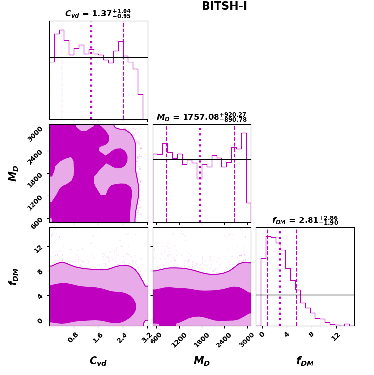}
    \caption{Similar to Fig. \ref{fig:corner_case1_1}, but for {\it Case II}.}
    \label{fig:corner_case2_1}
\end{figure*}

\begin{figure*}[t]
    \centering
    \includegraphics[width=0.52\linewidth]{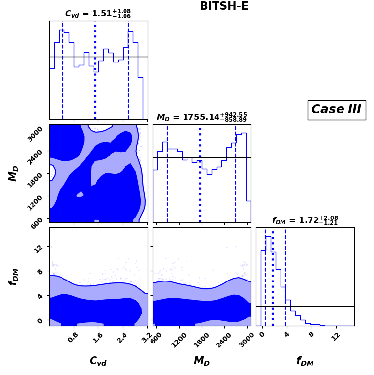}\includegraphics[width=0.52\linewidth]{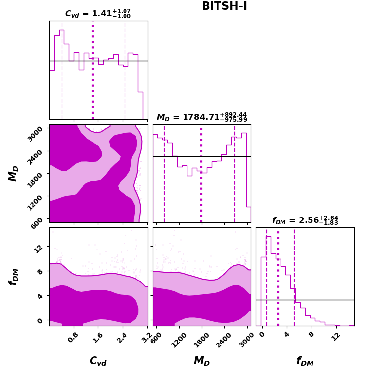}
    \caption{Similar to Fig. \ref{fig:corner_case1_1}, but for {\it Case III}.}
    \label{fig:corner_case3_1}
\end{figure*}

From Fig. \ref{fig:corner_case1_1}, we observe that 
when astrophysical constraints are imposed as per {\it Case I}, 
DM fraction ($f_{DM}$) in DMANS is well constrained as compared to the rest of the DM parameters {\it viz}., the coupling to the mass ratio, $C_{vd}$ and the mass of the DM particles, $M_D$ are poorly constrained. Note that this trend is observed for both BITSH-E and BITSH-I models. A slightly higher $f_{DM}$ of $\sim$ 3.44\%,  is obtained for the latter, which is stiffer of the two EoSs. The $C_{vd}$ parameter has a flat distribution with median values of 0.0154  MeV$^{-1}$ and 0.0161 MeV$^{-1}$ for BITSH-E and BITSH-I, respectively. Likewise, $M_D$ is also constrained to be $\sim$ 1993 MeV and 2050 MeV respectively, with a slightly heavier DM particle obtained for stiffer (BITSH-I) EoS. Similar trends in the constraining of $M_D$ are also reported in Ref. \cite{thakur2023exploring}.
\begin{figure*}[t]
    \centering
    \includegraphics[width=\linewidth]{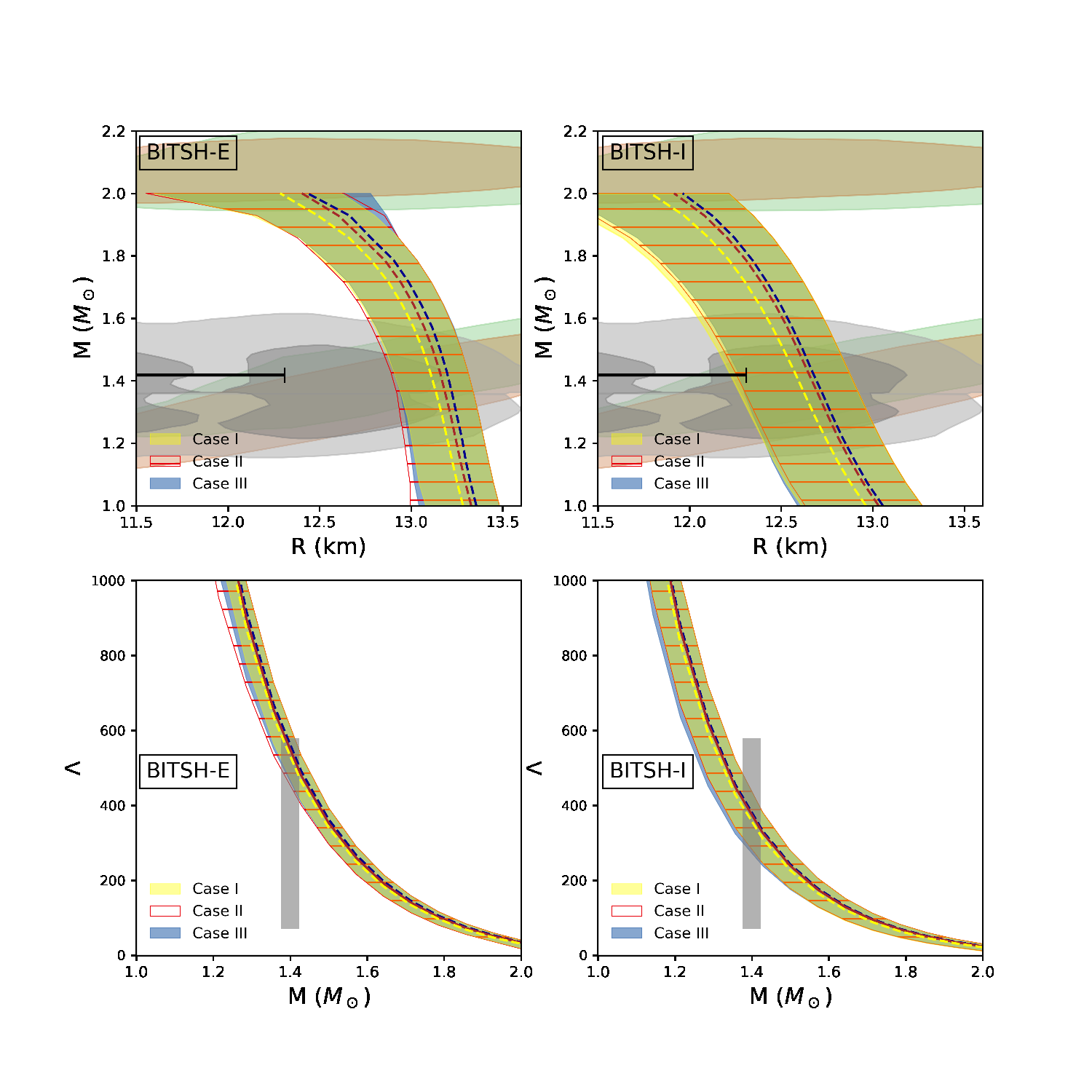}
    \caption{$M - R$ (upper panels) and $\Lambda - M$ (lower panels) distribution plots for BITSH-E (left) and BITSH-I (right) EoSs obtained with C $ \neq 0$
    %\SG{ C =0 OR non-zero, mention} 
    with medians (dashed lines) and the 95\% CI shown for {\it Cases I, II} and {\it III}. Medians and shaded regions follow the same color scheme. NICER X-ray observations are shown as Green (\cite{Miller_2019,Miller_2021}) and Orange (\cite{Riley_2019,Riley_2021}) contours for PSR J0740+6620 and PSR J0030+0451, together with the latest NICER measurements of mass and radius for PSR J0437-4715 (black error bar) \cite{Choudhury:2024xbk}. The grey shaded M-R regions represent the 90\% (light) and 50\% (dark) CI [top panels] and grey vertical band (90\% CI) represents the  tidal deformability [bottom panels] obtained from the LIGO/Virgo constraints derived from the binary components of GW170817 event.}
    \label{fig:MRposterior_all2}
\end{figure*}

Fig. \ref{fig:corner_case2_1} shows the posterior results when the constraints are imposed as per {\it Case II}, which incorporates the NICER M-R distributions through KDE.
However, we observe that only the $f_{DM}$ is well constrained and weak constraints are obtained on the $C_{vd}$ and the $M_D$. Here also, the stiffer EoS, BITSH-I, tends to acquire a larger DM fraction ($\sim$ 2.8\%) compared to BITSH-E ($\sim$ 2.3\%). These values are slightly lower than those seen earlier in {\it Case I}. 
$M_{DM}$, though not constrained well here, shows relatively lower median values as compared to those in {\it Case I}. 

As explained earlier, we expect more precise data with the next-generation detectors. Considering this, we generated mock data set with higher precision on mass and radius in the observations for pulsars PSR J740+6620 and PSR J0030+0451 and the astrophysical constraints are imposed following {\it Case III}. The posterior distributions of DM parameters obtained after incorporating the mock data into Bayesian analyses are shown in Fig. \ref{fig:corner_case3_1}. These results closely resemble those obtained from {\it Case II}. In particular, $f_{DM}$ is further reduced to approximately $\sim 1.7\%$ for BITSH-E and $\sim 2.6\%$ for BITSH-I EOS, compared to {\it Case II}. However, the other two DM parameters  $C_{vd}$ and $M_D$ remain poorly constrained. This suggests  $f_{DM}$ is the key parameter in determining the properties of DMANS, regardless of how astrophysical observations are incorporated into the likelihood function of the Bayesian analysis.
 
\noindent 
From the above analyses, we observe that BITSH-I EoS tends to have a higher DM fraction consistently in all three cases. We note that BITSH-I is somewhat softer compared to BITSH-E EoS at densities around 2-4$\rho_0$, but it becomes stiffer at higher densities. Consequently, the EoS stiffness at high densities seems to influence the fraction of DM in DMANS.

We further examine the properties of DMANS obtained by solving two-fluid TOV equations (Eq. 5). Fig. \ref{fig:MRposterior_all2} presents the distribution plots for the $M-R$ (upper panels) and $\Lambda-M$ (lower panels). The plots on the left (right) side correspond to the DMANS curves for BITSH-E (BITSH-I). These distributions are derived from the posterior samples obtained through Bayesian analysis for the three cases. The median $M-R$ curves (dashed lines) are shown along with the 95\% CI for each case. The CI bands are shown by different colors and hatched regions for the three cases in the figure ({\it Case I}: yellow, {\it Case II}: red, {\it Case III}: blue).

NICER X-ray observations are represented by green \cite{Miller_2019,Miller_2021} and orange contours \cite{Riley_2019,Riley_2021} for PSR J0030+0451 ($\sim$ 1.44M$_{\odot}$) and PSR J0740+6620 ($\sim$ 2.08M$_{\odot}$)  respectively, along with the latest NICER measurement of PSR J0437-4715 (black error bars) \cite{Choudhury:2024xbk}. The grey shaded regions represent the 90\% and 50\% CI of the LIGO/Virgo constraints derived from binary components of the GW170817 event \cite{Abbott_2018}. Furthermore, the observational constraints (90\% CI) on the tidal deformability of the same event are shown as a vertical band in the bottom panel. The similarity in the spread of the distributions in both $M-R$ and $\Lambda-M$ plots suggests that various methods of applying astrophysical constraints have minimal impact on the properties of DMANS.

\begin{figure*}[t]
    \centering
    \includegraphics[width=\linewidth]{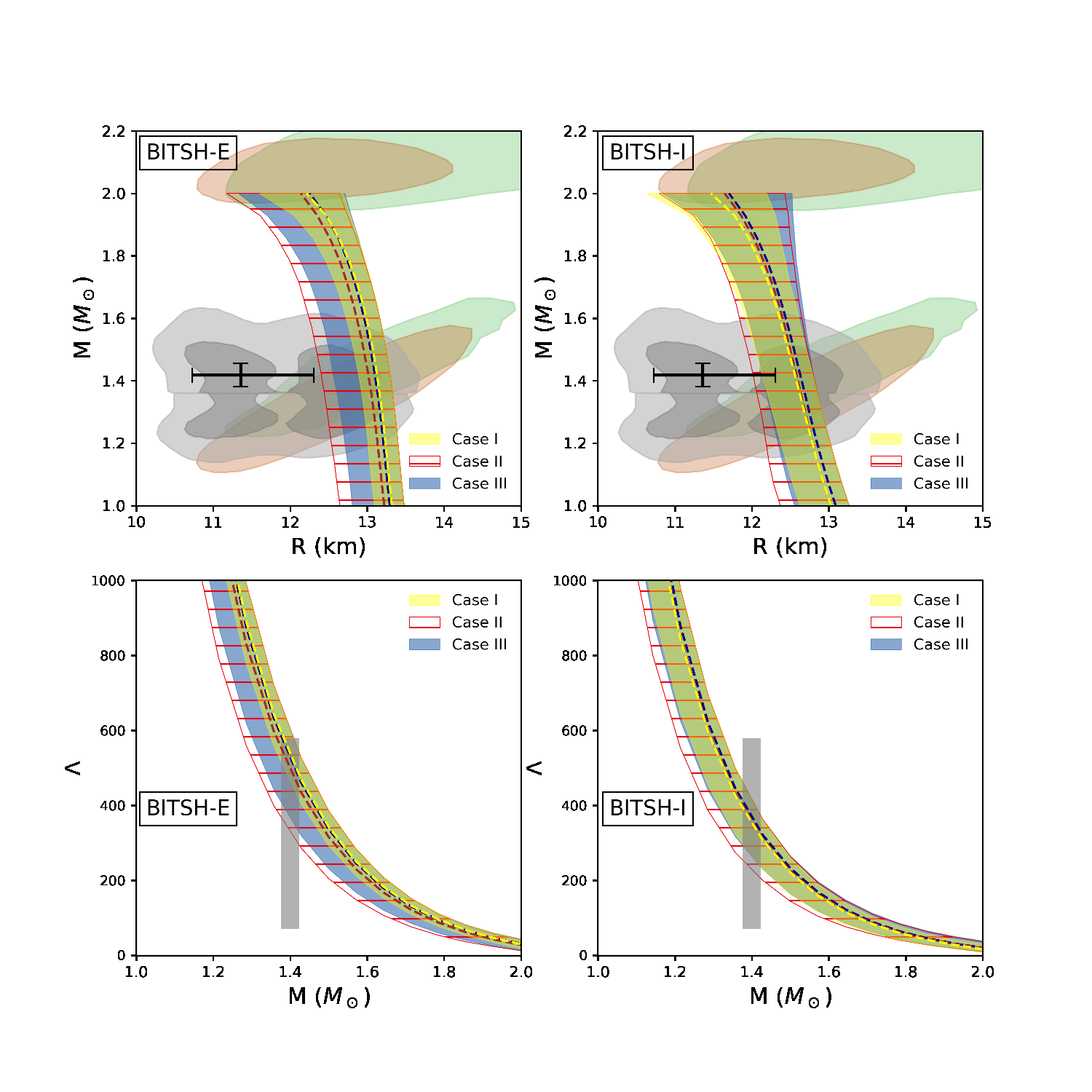}
    \caption{ Same as Fig. \ref{fig:MRposterior_all2}, for models with C $\neq$ 0 but when high-density EoS is allowed to vary while modeling DM parameters.}
    \label{fig:MRposterior_all1}
\end{figure*}

\subsection{Role of High Density Uncertainties (HDU) in HM EoS in constraining DM parameters}\label{HM-HDU}

Furthermore, we investigate the role of high density uncertainties in the EoS in governing the DM parameters. The internal composition of NS at high densities ($\rho > 2 \rho_0$) is still not well known and the appearance of exotic matter with new degrees of freedom is one of the plausible explanations for this uncertainty. Therefore, sometimes EoSs beyond the transition density (1.5-2$\rho_0$) are simply constructed through the speed of sound parametrization as explained in Sec \ref{HD EoS}. Here, the Bayesian analysis incorporates both the DM parameters and those for the high density speed of sound parametrization.

As a result, the posteriors are generated for the two models BITSH-E and BITSH-I and in each case, we vary the EoS at high density to allow  for the above uncertainty as per Eq. \ref{cs2}. We analyzed the posterior distributions obtained with the above high-density uncertainties (HDU) and the median values of the posteriors obtained for the DM parameters are listed in Table \ref{table:posteriors} under Section ``HM EoS with HDU". From the table, it is seen that $C_{vd}$ and $M_D$ are again not well constrained, but the $f_{DM}$ appears to be relatively well constrained, as observed in earlier analyses. For {\it Case I}, a higher DM fraction of $\sim$ 3.6\% is observed in BITSH-I EoS in contrast to $\sim$ 2.4\% in BITSH-E EoS.  
However, for {\it Case II} and {\it Case III}, the trend reverses and BITSH-E EoS prefers to have higher dark matter. This can be understood as follows. When the uncertainties at high-density EoSs are considered, the difference in the stiffness of the two EoSs becomes insignificant at such densities. Moreover, BITSH-I EoS is softer than BITSH-E EoS at densities relevant to the constraints of a canonical mass star (used in {\it Case II} and {\it Case III}), thereby leading to a higher DM fraction in stiffer BITSH-E EoS. 

Furthermore, we analyze the properties of neutron stars derived from the HM EoS combined with HDU. 
We display the distribution curves $M-R$ and $\Lambda-M$ for two models with HD variations in Fig. \ref{fig:MRposterior_all1}. Here, the spread of the posterior distributions appears to depend on the way the constraints are imposed. A relatively narrower spread is observed for {\it Case I } (yellow) both in $M-R$ and $\Lambda-M$ plots. Moreover, {\it Case II} (red) has a wider distribution than {\it Case III} (blue),  due to the higher uncertainties in the mass-radius data sets in the former case. 
This effect is more pronounced in the BITSH-E EoS, yet the medians remain largely unaffected by the application of astrophysical constraints.

%\newgeometry{margin=1.2cm}
\begin{table*}[t]
    \centering
    \caption{Posteriors obtained for the DM parameters; $C_{vd}, M_D$ and $f_{DM}$ after Bayesian analyses. The three subsections of the table correspond to the HM EoS fixed over entire density range, HM EoSs with HD uncertainty and stiffer HM EoSs without $\omega^4$ term (C = 0). The medians values ($\pm 1 \sigma$) are listed for {\it Case I, II} and {\it III}.} 
    \vspace{0.5cm}
    \renewcommand{\arraystretch}{2.1}
    \begin{tabular}{|*{7}{>{\centering\arraybackslash}p{0.13\linewidth}|}} %{|l|ccc |ccc|}
    \hline\hline 
    %\centering
    %\vspace{0.01cm}
    \multicolumn{1}{|l|}{\textbf{Models}}  & \multicolumn{3}{c|}{\textbf{BITSH-E}} & \multicolumn{3}{c|}{\textbf{BITSH-I}} \\ 
    \hline 
    DM parameter & $C_{vd}$  (MeV$^{-1})$ & $M_D$ (MeV) &  $f_{DM}$ (\%) & $C_{vd}$  (MeV$^{-1})$ & $M_D$ (MeV) &  $f_{DM}$ (\%)\\
     & \small{(x$10^{-2}$)} & & & \small{(x$10^{-2}$)} & & \\
    \hline\hline  
    \multicolumn{7}{|c|}{\textbf{Fixed HM EoS}}\\
    \hline   
    {\it Case I} & $1.54^{+1.04}_{-1.11}$ & $1992.83^{+1100.91}_{-984.97}$ & $2.46^{+1.49}_{-1.65}$ & $1.61^{+0.98}_{-1.15}$ & $2049.96^{+973.43}_{-983.56}$ & $3.44^{+2.46}_{-2.17}$\\
    \hline 
    {\it Case II} & $1.52^{+1.04}_{-1.03}$ & $1739.25^{+979.92}_{-862.07}$ & $2.28^{+2.63}_{-1.57}$ & $1.37^{+1.04}_{-0.95}$ & $1757.07^{+920.26}_{-890.77}$ & $2.81^{+2.86}_{-1.90}$\\
    \hline
    {\it Case III} & $1.51^{+1.08}_{-1.06}$ & $1755.13^{+942.55}_{-858.89}$ & $1.72^{+2.08}_{-1.21}$ & $1.41^{+1.07}_{-1.00}$ & $1784.70^{+892.43}_{-975.98}$ & $2.56^{+2.64}_{-1.83}$\\
    \hline\hline
    \multicolumn{7}{|c|}{\textbf{HM EoS with HDU}}\\
    \hline
    {\it Case I} & $1.53^{+0.97}_{-1.04}$ & $1962.62^{+1069}_{-938.55}$ & $2.36^{+1.38}_{-1.52}$ & $1.48^{+0.97}_{-0.96}$ & $2099.68^{+935.17}_{-986.67}$ & $3.65^{+2.22}_{-2.28}$\\
    \hline
    {\it Case II} & $1.50^{+0.95}_{-0.94}$ & $1706.20^{+836.3}_{-740.55}$ & $3.36^{+3.37}_{-2.19}$ & $1.48^{+0.97}_{-0.99}$ & $1832.88^{+767.58}_{-836.21}$ & $2.85^{+3.11}_{-1.91}$\\
    \hline
    {\it Case III} & $1.49^{+0.94}_{-1.00}$ & $1849.82^{+774.37}_{-888.29}$ & $2.40^{+2.73}_{-1.57}$ & $1.43^{+1.06}_{-0.95}$ & $1818.82^{+779.63}_{-846.20}$ & $2.28^{+2.75}_{-1.57}$\\
    \hline\hline 
    \multicolumn{7}{|c|}{\textbf{HM EoS (C = 0) with HDU}}\\
    \hline 
    {\it Case I} & $1.58^{+0.91}_{-0.96}$ & $1936.31^{+939.70}_{-812.02}$ & $8.12^{+5.74}_{-4.86}$ & $1.57^{+0.94}_{-1.01}$ & $2056.07^{+968.79}_{-972.54}$ & $4.37^{+3.52}_{-2.83}$\\
    \hline 
    {\it Case II} & $1.58^{+0.93}_{-0.95}$ & $1783.23^{+826.67}_{-762.95}$ & $4.82^{+4.73}_{-2.97}$ & $1.54^{+0.93}_{-1.04}$ & $1833.26^{+834.36}_{-903.34}$ & $2.92^{+3.51}_{-1.99}$\\
    \hline 
    {\it Case III} & $1.53^{+0.95}_{-1.01}$ & $1689.08^{+858.60}_{-782.98}$ & $3.58^{+4.00}_{-2.41}$ & $1.58^{+0.92}_{-1.10}$ & $1820.58^{+800.18}_{-883.46}$ & $2.35^{+2.92}_{-1.63}$\\
    \hline
    \end{tabular}
    \label{table:posteriors}
\end{table*}		
%\restoregeometry

\subsection{Role of stiffness of HM EoS in constraining DM parameters}\label{HM-stiff}

To ensure the robustness of our findings, we repeat the Bayesian analyses using the HM Lagrangian without the $\omega^4$ term (C = 0 in Eq.\ref{eq:lag}), which results in significantly stiffer EoSs. This analysis also includes the high-density uncertainty for both the EoSs. We obtain similar posterior distributions for all three cases, where only the DM fraction exhibits a narrow posterior distribution and is relatively better constrained compared to the other two DM parameters ($C_{vd}$ and $M_D$).  Note that DM fraction is  a crucial parameter that  determines the amount of DM in NS for different central energy densities. Moreover, it is directly related to the total mass of DM in NS, influencing the mass-radius of the admixed NS.  Consequently it is physically consistent that $f_{DM}$  is the  most constrained parameter in the present analyses. Note that we also observe that $M_{max}$, $R_{1.4}$ and $\Lambda_{1.4}$ of DMANS exhibit a relatively strong correlation  with $f_{DM}$, whereas no significant correlation is found with $C_{vd}$ and $M_{DM}$.

We observe that when the Lagrangian excludes the $\omega^4$ interaction term, the BITSH-E EoS consistently yields a higher DM fraction, across all the cases. For example, in {\it Case I} ({\it Case II}), the DM fraction is approximately $\sim 8.1\% (4.8\%)$ in BITSH-E, compared to $4.4\% (2.9\%)$ in BITSH-I. This is because BITSH-E remains stiffer than BITSH-I across all densities when the $\omega^4$ term is omitted. In particular, this contrast between the EoS BITSH-E and BITSH-I was not evident earlier (when C $\neq$ 0), as their stiffness at high density did not differ significantly. The maximum neutron star masses obtained for BITSH-E and BITSH-I EoSs (with C $\neq$ 0) are $2.11 M_{\odot}$ and $2.18 M_{\odot}$, respectively. However, when C = 0, the difference becomes more pronounced, with maximum masses increasing to approximately $2.51 M_{\odot}$ for BITSH-E and $2.26 M_{\odot}$ for BITSH-I EoS.

Furthermore, the median value of $C_{vd}$ remains around $0.015  \text{MeV}^{-1}$ in all cases for both EoSs. The DM particle mass is consistently higher for BITSH-I compared to BITSH-E. The posterior distributions of the three DM parameters for {\it Cases I, II}, and {\it III} (without the $\omega^4$ term) are summarized in Table \ref{table:posteriors}. We would like to mention that similar results have been observed for bosonic dark matter, where the fraction of DM mass within NS is constrained by mass-radius measurements. However, despite the stringent constraints on the baryonic equation of state, the mass and coupling constant of dark matter remain largely unconstrained \cite{Rutherford_2023}.

For all EoSs considered in this study, our comprehensive analysis reveals that, in {\it Case II} scenario-where astrophysical constraints are treated in a more sophisticated manner—the median value of the posteriors obtained for DM fraction averages around 3\% across different EoS regimes. The corresponding maximum fraction at the 2$\sigma$ confidence level is approximately 14.2\%.

We would also like to highlight that, in {\it Case II}, the peaks of the posterior distributions lie below 2\% across the different EoSs. This suggests that smaller dark matter fractions are more probable, although higher fractions—up to about 10\%—remain permissible.   Similar results are reported in Ref. \cite{Rutherford_2023,thakur2023exploring},
where smaller DM fractions below 10\% are seen in posterior distributions. Furthermore, 4$\sigma$ deviations in the posteriors are reported to be around 14\% \& 19 \% for different scenarios of NS mass-radius measurements in Ref. \cite{Rutherford_2023}, which are also consistent with those obtained in the present study.

Furthermore, to explore the broader effect of hadronic equations of state on DM parameters, we extended our analysis to four additional EoSs corresponding to distinct interaction terms in the Lagrangian as in Eqn. \ref{eq:lag}. Specifically, we repeat the above analysis for NL3 \& GM1 (where C = 0, $\Lambda_v$ = 0) and MS1 \& TM1 (where C $\ne$ 0, $\Lambda_v$ = 0). The posteriors obtained for these EoSs along with the previous ones, BITSH-E \& BITSH-I are listed in Table \ref{table:4newmodels}. These analyses are also performed with the HDU as discussed in section \ref{HM-HDU} and constraints are imposed as in {\it Case II}. The largest median DM fraction, of approximately 15\%, is observed for NL3 EoS, consistent with its characteristically high stiffness. In contrast, the remaining EoSs yield lower DM fractions, with median values below 5\%. Notably, the posterior medians for the DM coupling constant and particle mass remain of the same order across all EoSs considered. This analysis further validates that the DM fraction is predominantly governed by the stiffness of the HM EoS. It is worth emphasizing that the HM EoS was not varied within the Bayesian inference framework by adopting a broader prior distribution of nuclear matter parameters. This choice was made to avoid computationally expensive analyses and introducing additional uncertainties from the hadronic sector, which would further hinder the ability to place meaningful constraints on the DM parameter space. Such degeneracies have been observed in a previous study \cite{Rutherford_2023}, where DM particles were modeled to be bosonic in nature.
Instead, we have allowed the EoS parameters at high densities to be modeled in Bayesian analyses, to take care of the uncertainties in EoS beyond 2$\rho_0$.

\begin{table*}
    \centering
    \caption{Posteriors for the DM parameters for the given HM models \cite{SunAndLattimer2024}. The constraints are imposed as in {\it Case II} with HDU. The results for models, BITSH-E and BITSH-I are displayed again for completeness. }
    \renewcommand{\arraystretch}{2}
    \begin{tabular}{|*{5}{>{\centering\arraybackslash}p{0.16\linewidth}|}} %|c|c|c|c|}
    \hline
    \textbf{Interaction terms} & \textbf{Models} & $C_{vd}$  (MeV$^{-1})$ & $M_D$ (MeV) &  $f_{DM}$ (\%)\\ 
    & & \small{(x$10^{-2}$)} & & \\
    \hline
    \multirow{2}{*}{C=0, $\Lambda_v$=0} & \textbf{GM1} & $1.63^{+0.85}_{-0.92}$ & $1835.18^{+732.33}_{-787.64}$ & $5.34^{+4.12}_{-3.28}$\\\cline{2-5}
     & \textbf{NL3} & $1.41^{+0.89}_{-0.76}$ & $1841.52^{+697.88}_{-700.78}$ & $15.39^{+3.08}_{-4.38}$\\
    \hline\hline
    \multirow{2}{*}{C$\ne$0, $\Lambda_v$=0} & \textbf{MS1} & $1.71^{+0.84}_{-1.02}$ & $1704.71^{+759.38}_{-704.62}$ & $5.63^{+4.89}_{-2.95}$\\\cline{2-5}
     & \textbf{TM1} & $1.69^{+0.86}_{-0.94}$ & $1580.95^{+865.59}_{-668.57}$ & $5.19^{+3.87}_{-2.68}$\\
    \hline
    \hline
    \multirow{2}{*}{C$\ne$0, $\Lambda_v \ne$0} & \textbf{BITSH-E} & $1.50^{+0.95}_{-0.94}$ & $1706.20^{+836.3}_{-740.55}$ & $3.36^{+3.37}_{-2.19}$ \\\cline{2-5}
     & \textbf{BITSH-I} & $1.48^{+0.97}_{-0.99}$ & $1832.88^{+767.58}_{-836.21}$ & $2.85^{+3.11}_{-1.91}$\\
    \hline
    \hline
    \multirow{2}{*}{C=0, $\Lambda_v \ne$0} & \textbf{BITSH-E} & $1.58^{+0.93}_{-0.95}$ & $1783.23^{+826.67}_{-762.95}$ & $4.82^{+4.73}_{-2.97}$ \\\cline{2-5}
     & \textbf{BITSH-I} & $1.54^{+0.93}_{-1.04}$ & $1833.26^{+834.36}_{-903.34}$ & $2.92^{+3.51}_{-1.99}$\\
     \hline
    \end{tabular}
    \label{table:4newmodels}
\end{table*}

\section{Summary}\label{summary}
We have explored the properties of dark matter admixed-neutron stars (DMANS) by considering fermionic dark matter which is confined to the core of NS and is interacting via gravitational interaction with hadronic matter. We have employed equations of state (EoSs) for the hadronic sector and dark sector within a relativistic mean-field approach and two-fluid TOV equations are solved to obtain NS properties. A set of realistic equations of state for the hadronic sector are used which are recently derived based on the properties of finite nuclei, experimental data from heavy-ion collisions and observations of mass, radius and tidal deformability of neutron stars. The parameters of the dark matter (DM) model, {\it viz.} DM particle mass, DM mass fraction and coupling parameter, are constrained using a Bayesian inference technique using different likelihood functions for the treatment of astrophysical observations of NSs. Whereas, in one case, constraints of maximum mass and radii of low-mass stars are imposed; the other one has involved complete mass-radius distributions of NICER for PSR J0740+6620 and PSR J0030+0451. A third scenario is also considered where a mock data set of NICER observations with reduced uncertainties is considered in the likelihood function, anticipating much precised future observations. The above analyses are also carried out by varying the high-density hadronic matter EoSs within the conformal limit. Additionally, we assess the robustness of our findings by conducting the study with comparatively stiffer EoSs in the hadronic sector, where the self-interaction term of vector mesons is ignored. It is observed that in all three scenarios, DM fraction is well constrained and rest of the two parameters are poorly constrained. The DM mass-fraction is the most constrained parameter in this analysis because it directly influences the total mass of DM in the NS, which in turn affects the star's mass and radius. To broaden our scope of study, we extended the analysis for other EoSs of hadronic matter, where different interaction terms in the Lagrangian are considered. DM fraction inside a NS is seen to be determined by the stiffness of the EoS, where stiffer hadronic EoS tends to acquire a higher DM fraction relative to its softer counterpart. Based on our detailed analyses on DM parameters obtained for different HM Lagrangians and with high-density EoSs, referring to Tables \ref{table:posteriors} and \ref{table:4newmodels} we can summarize our findings as:
\begin{enumerate}
    \item[$\ast$] Current astrophysical constraints are limited to restricting the DM fraction in DMANS, without providing any information on the mass of the DM particle or its coupling parameters.
    \item[$\ast$] The DM fraction is mostly unaffected by how astrophysical observations are integrated in the analyses.
    \item[$\ast$] The uncertainty in the high-density EoS has weak impact on the DM fraction.
    \item[$\ast$] The DM fraction is primarily determined by the stiffness of the HM EoS at high densities.
\end{enumerate}
Therefore, we can assert that the DM fraction is the primary factor driving the properties of DMANS, regardless of how current astrophysical observations are integrated into the statistical inference method. The average DM mass fraction obtained for realistic EoSs that aligns with well-established astrophysical observations lies below 5\%. We can uncover more mysteries of DM using other EoSs that include exotic matter in the NS core, along with more advanced astrophysical observations expected in the near future.

\section{Acknowledgment}
PA acknowledges the computing facilities at BITS Pilani-Hyderabad Campus, used for the analyses. PA also wishes to thank D. G. Roy for the fruitful interactions. AV acknowledges the CSIR-HRDG for support through the CSIR-JRF 09/1026(16303)/2023-EMR-I.

%\newpage 

\bibliographystyle{apsrev4-2}
\bibliography{DM_NS}
\end{document}